\documentclass[prc,aps,a4paper,groupedaddress,superscriptaddress,nofootinbib,showpacs
,preprintnumbers,onecolumn]{revtex4}
\usepackage{graphicx}% Include figure files
\usepackage{amsfonts}
\usepackage{amssymb}
\usepackage{natbib}
\usepackage{dcolumn}% Align table columns on decimal point
\usepackage{bm}% bold math
\newcommand{\bwt}{\begin{widetext}}
\newcommand{\ewt}{\end{widetext}}
\newcommand{\beq}{\begin{equation}}
\newcommand{\eeq}{\end{equation}}
\newcommand{\bea}{\begin{eqnarray}}
\newcommand{\eea}{\end{eqnarray}}
\begin{document}
\title{Comment on "Energies of the ground state and first excited $0^{+}$ 
state in an exactly solvable pairing model"}
\author{A. Rabhi}
\email{rabhi@ipnl.in2p3.fr}
\affiliation{Laboratoire de Physique de la Mati\`ere Condens\'ee, 
Facult\'e des Sciences de Tunis, Campus Universitaire, Le Belv\'ed\`ere-1060, 
Tunisia} 
\affiliation{IPN-Lyon, 43Bd du 11 novembre 1918, F-69622 Villeurbanne Cedex, 
France}
\date{\today}
\begin{abstract}
We comment on a recent application of the RPA method and its extensions to the 
case of the two-level pairing model by N.~Dinh Dang~\cite{b1}. 
\end{abstract}
\maketitle
\begin{figure}[htb]                       
\centering
\vspace{0.3 in}
\includegraphics[width=0.5\linewidth]{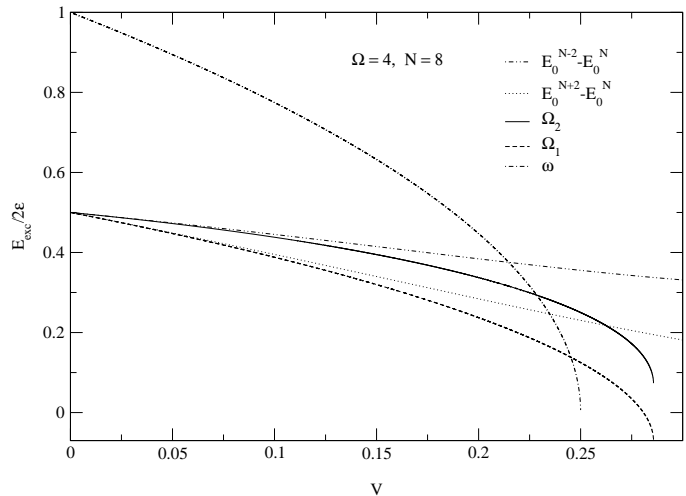} 
\caption{Excitation energies $E_{exc}$ in the non superfluid region as function of 
$V=\Omega/2\epsilon$ described in the text, for particle number $N=8$. The spin of 
the levels is $J=7/2$. 
The results refer to to exact calculations $2\mu^{\pm}=\pm (E^{N\pm 2}_{0}-E^{N}_{0})$ 
(double-dot dashed line, dotted line), standard pp-RPA Eqs.(\ref{c7}) (solid line, dashed line) 
and pp-RPA in the single boson approximation Eq.(\ref{c9}) (dot-dashed line).}
\label{eigen}
\end{figure}

The aim of this comment is to discuss and explain several aspects related to 
the derivation and application of the RPA theory in relation with
a recent paper~\cite{b1} of Dinh Dang, there RPA theory has been applied 
to the two-level pairing model. We have some criticisms and remarks. 
We will start with the RPA in the boson formalism as used by N.~Dinh Dang~\cite{b1}.
We notice that Dinh Dang uses a quite unconventional and non systematic boson expansion.
Let us restate the standard boson mapping in the case of the two level pairing model
\beq
H=\frac{\epsilon}{2}(\hat{N}_{2}-\hat{N}_{1}) - g\Omega \sum_{j
{j}^{\prime}} A_{j}^{\dagger} A_{{j}^{\prime}}, \; j,{j}^{\prime} =1,2.
\label{a1}
\eeq
To lowest order one have
\bea
A^{\dagger}_{1}=\frac{1}{\sqrt{\Omega}}\sum_{m>0}a^{\dagger}_{1m}a^{\dagger}_{1\bar{m}}&\rightarrow& b_{1}, \cr
A^{\dagger}_{2}=\frac{1}{\sqrt{\Omega}}\sum_{m>0}a^{\dagger}_{2m}a^{\dagger}_{2\bar{m}}&\rightarrow& b^{\dagger}_{2} 
\label{c1}
\eea
where $b^{\dagger}_{i}$, $b_{i}$ are ideal Bose operators. The occupation number operators obey 
the exact relations
\bea
\hat{N}_{1}=\sum_{m}a^{\dagger}_{1m}a_{1m}&\rightarrow& 2 (\Omega - b^{\dagger}_{1}b_{1})\cr
\hat{N}_{2}=\sum_{m}a^{\dagger}_{2m}a_{2m}&\rightarrow& 2 b^{\dagger}_{2}b_{2}.
\label{c2}
\eea
Dinh Dang now uses the particle number condition  
$\langle\hat{N}_{1}\rangle +\langle\hat{N}_{2}\rangle=\langle\hat{N}\rangle\equiv N\equiv
2\Omega$, which holds if, in the absence of interaction, the lowest level is filled. 
From the above number condition, one obtains
\beq
\langle b^{\dagger}_{1}b_{1}\rangle=\langle b^{\dagger}_{2}b_{2}\rangle.
\label{c3}
\eeq
Dinh Dang deduces from this relation that $b_{1}=b_{2}=b, \quad b^{\dagger}_{1}=b^{\dagger}_{2}=b^{\dagger}$,
\textit{i.e.} it is assumed that it is approximately valid to replace the two ideal bosons $b_{1}$ 
and $b_{2}$ by the single one $b$. In the first part of this work we want to study the validity 
of this single boson approximation. Keeping the two bosons the pairing Hamiltonian is given to 
lowest order by 
\bea
H&=&-\frac{\epsilon}{2}2(\Omega - b^{\dagger}_{1}b_{1})+\frac{\epsilon}{2}2
b^{\dagger}_{2}b_{2}-g\Omega\sum_{i,j=1,2}b^{\dagger}_{i}b_{j} \cr
&=&-\epsilon(\Omega - b^{\dagger}_{1}b_{1})+\epsilon b^{\dagger}_{2}b_{2}
-g\Omega(b^{\dagger}_{2}b_{2}+b_{1}b^{\dagger}_{1} \cr
&+&b^{\dagger}_{2}b^{\dagger}_{1}+b_{1}b_{2}).
\label{c4}
\eea
In the single boson approximation we have 
\bea
H&=&-\frac{\epsilon}{2}2(\Omega -b^{\dagger}b)+\frac{\epsilon}{2}2
b^{\dagger}b-g\Omega(b^{\dagger}b+bb^{\dagger}+b^{\dagger}b^{\dagger}+bb) \cr
&=&-\epsilon(\Omega - b^{\dagger}b)+\epsilon b^{\dagger}b
-g\Omega(b^{\dagger}b+bb^{\dagger}+b^{\dagger}b^{\dagger}+bb).
\label{c5}
\eea
Both Hamiltonians can trivially be diagonalized with the help of an RPA (Bogoliubov transformation) 
among the bosons. We obtain for~(\ref{c4})
\beq
H=\Omega_{1}b^{\dagger}_{1}b_{1}+\Omega_{2}b^{\dagger}_{2}b_{2}
\label{c6}
\eeq
with,
\bea
\Omega_{1}&=&-g+\sqrt{g+\epsilon}\sqrt{\epsilon+g(1-2\Omega)} \cr
\Omega_{2}&=& g+\sqrt{g+\epsilon}\sqrt{\epsilon+g(1-2\Omega)}
\label{c7}
\eea
and for~(\ref{c5})
\beq
H=\omega b^{\dagger}b
\label{c8}
\eeq
with,
\beq
\omega = 2\epsilon\sqrt{1-\frac{2g\Omega}{\epsilon}}.
\label{c9}
\eeq 
In Fig.~\ref{eigen} we have traces these different eigenvalues as a function of $V=g\Omega/2\epsilon$. 
Also shown are the exact values for the chemical potentials 
$2\mu^{\pm}=\pm (E^{N\pm 2}_{0}-E^{N}_{0})$ where $E^{N}_{0}$ are the exact 
groundstate energies obtained by diagonalization of the original pairing Hamiltonian.
The reason why we compare with $2\mu^{\pm}$ is given by the fact that the eigenvalues 
of standard pp-RPA have to be identified with these chemical potentials (see \textit{e.g.}
\cite{b4,b3,b2}). 
We also see that the eigenvalues $\Omega_{1}$, $\Omega_{2}$ follow, at least 
for small values of $V$, quite closely the exact values. On the other 
hand the single boson approximation yields a completely erroneous result which seems to 
have nothing to do with the exact solution. 

Let us now comment in the superfluid phase. We note that in the superfluid phase we can calculate
in standard QRPA the contribution of the q-term (see Eq.~(7) in~\cite{b1}) of the Hamiltonian. 
Consequently, there is no reason to neglect this term as discussed by N.~Dinh~Dang~\cite{b1}. 
The standard QRPA matrices are well-known and given by 
\bea
A_{jj^{\prime}}&=&2\left(E_{j}+2q_{jj^{\prime}}\right)\delta_{jj^{\prime}}+d_{jj^{\prime}} \\
B_{jj^{\prime}}&=&2\left(1-\frac{1}{\Omega}\delta_{jj^{\prime}}\right)h_{jj^{\prime}}
\label{c10}
\eea
where the quasiparticle energy is
$E_{j}=\sqrt{\left(\epsilon_{j}-gv^{2}_{j}-\mu\right)^{2}+\Delta^{2}}$ and the gap $\Delta$, 
as calculated from the BCS equation, includes the self-energy term. We have used the same 
notation as indicated in the work of Dinh Dang. We remind, shortly, that in this case, 
the gap is given by
\beq
\Delta=\sqrt{g^{2}\Omega^{2}-\frac{\xi^{2}}{4}},
\label{c11}
\eeq
together with
\bea
u^{2}_{1}=v^{2}_{2}=\frac{1}{2}\left(1-\frac{\xi}{2g\Omega}\right), \\
v^{2}_{1}=u^{2}_{2}=\frac{1}{2}\left(1+\frac{\xi}{2g\Omega}\right),
\label{c12}
\eea  
\beq
\mu=-\frac{g}{2},
\label{c13}
\eeq
where $\xi$ is defined as $\xi=2\epsilon\Omega/(2\Omega-1)$. 
Using the latter relations we can write
\bea
u^{2}_{2}-v^{2}_{2}&=&\frac{\xi}{2g\Omega};
u_{1}v_{1}u_{2}v_{2}=u^{2}_{2}v^{2}_{2}=\frac{\Delta^{2}}{4g^{2}\Omega^{2}}\\
u^{4}_{2}+v^{4}_{2}&=&\frac{1}{2}+\frac{\xi^{2}}{8g^{2}\Omega^{2}};
v^{4}_{2}=\frac{1}{4}+\frac{\xi^{2}}{16g^{2}\Omega^{2}}-\frac{\xi}{4g\Omega} 
\label{c14}
\eea 
and we can calculate the different contribution of each term $E_{j}$, $q_{jj^{\prime}}$, $d_{jj^{\prime}}$, 
and $h_{jj^{\prime}}$
\bea
E_{1}=E_{2}=g\Omega +\frac{\Delta^{2}}{4g\Omega^{2}};
q_{11}=q_{22}=-\frac{\Delta^{2}}{4g\Omega^{2}}\\
d_{11}=d_{22}=-g\Omega +\frac{\Delta^{2}}{2g\Omega};
d_{12}=d_{21}=-\frac{\Delta^{2}}{2g\Omega} \\
h_{11}=h_{22}=\frac{\Delta^{2}}{4g\Omega};
h_{12}=h_{21}=\frac{g\Omega}{2}-\frac{\Delta^{2}}{4g\Omega}.
\label{c15}
\eea
Therefore, explicitly, the matrix elements are given by
\bea
A_{11}&=&A_{22}=g\Omega -\frac{\Delta^{2}}{2g\Omega^{2}}+\frac{\Delta^{2}}{2g\Omega} \\
A_{12}&=&A_{21}= -\frac{\Delta^{2}}{2g\Omega}\\
B_{11}&=&B_{22}=-\frac{\Delta^{2}}{2g\Omega^{2}}+\frac{\Delta^{2}}{2g\Omega} \\
B_{12}&=&B_{21}= g\Omega - \frac{\Delta^{2}}{2g\Omega}
\label{c16}
\eea
and the positive eigenvalues of the RPA matrix are given by
\beq
\Omega_{1}=0, \quad\Omega_{2}=2\Delta\sqrt{1-\frac{1}{2\Omega}}
\label{c17}
\eeq
in agreement with the result found by \textit{e.g.} Hagino and Bertsch~\cite{b3}.
In the case of the Fermion formalism of Dinh Dang, to obtain the RPA matrix elements, we have 
to begin with (see Eqs.~(57-61) in~\cite{b1})
\bea
A_{jj^{\prime}}&=&2\left(E_{j}+3q_{jj^{\prime}}\right)\delta_{jj^{\prime}}+d_{jj^{\prime}} \cr
B_{jj^{\prime}}&=&2\left(1-\frac{1}{\Omega}\delta_{jj^{\prime}}\right)h_{jj^{\prime}}.
\label{c18}
\eea
where the factor $3$ is different from the correct factor appearing in~(\ref{c10}).
Explicitly, in the case~(\ref{c18}), the QRPA matrices are given, as follows
\bea
A_{11}&=&A_{22}=g\Omega -\frac{\Delta^{2}}{g\Omega^{2}}+\frac{\Delta^{2}}{2g\Omega} \\
A_{12}&=&A_{21}=-\frac{\Delta^{2}}{2g\Omega} \\
B_{11}&=&B_{22}=-\frac{\Delta^{2}}{2g\Omega^{2}}+\frac{\Delta^{2}}{2g\Omega} \\
B_{12}&=&B_{21}=g\Omega -\frac{\Delta^{2}}{2g\Omega}
\eea
where the gap $\Delta$ as calculated from the BCS equation includes the self-energy
term. The RPA eigenvalues are given by (see Eq.(65) in~\cite{b1})
\beq
\Omega_{1}=\Delta\sqrt{\frac{1}{\Omega}\left(\frac{3\Delta^{2}}{4g^{2}\Omega^{3}}-1\right)}
\eeq
\beq
\Omega_{2}=2\Delta\sqrt{\left(1-\frac{3}{4\Omega}\right)
\left(1-\frac{\Delta^{2}}{4g^{2}\Omega^{3}}\right)}.
\eeq
When we neglect the q-term (see Eq.~(7) in~\cite{b1}), we obtain
\bea
A_{11}&=&A_{22}=g\Omega +\frac{\Delta^{2}}{2g\Omega^{2}}+\frac{\Delta^{2}}{2g\Omega} \\
A_{12}&=&A_{21}=-\frac{\Delta^{2}}{2g\Omega}\\
B_{11}&=&B_{22}=-\frac{\Delta^{2}}{2g\Omega^{2}}+\frac{\Delta^{2}}{2g\Omega} \\
B_{12}&=&B_{21}=g\Omega -\frac{\Delta^{2}}{2g\Omega}
\eea
\beq
\Omega_{1}=\Delta\sqrt{\frac{2}{\Omega}}, \quad \Omega_{2}=2\Delta\sqrt{1+\frac{\Delta^{2}}{2g^{2}\Omega^{3}}}
\eeq
what leads to different RPA eigenvalues of that given by Dinh Dang for this case. 
However, if we divide the q-term by a factor of 2, we obtain
\bea
A_{11}&=&A_{22}=g\Omega +\frac{\Delta^{2}}{2g\Omega}\\
A_{12}&=&A_{21}=-\frac{\Delta^{2}}{2g\Omega}\\
B_{11}&=&B_{22}=-\frac{\Delta^{2}}{2g\Omega^{2}}+\frac{\Delta^{2}}{2g\Omega} \\
B_{12}&=&B_{21}=g\Omega -\frac{\Delta^{2}}{2g\Omega}
\eea
which produces the following RPA eigenvalues
\beq
\Omega_{1}=\Delta\sqrt{\frac{1}{\Omega}\left(1-\frac{\Delta^{2}}{4g^{2}\Omega^{3}}\right)}
\eeq
\beq
\Omega_{2}=2\Delta\sqrt{\left(1-\frac{1}{4\Omega}\right)
\left(1+\frac{\Delta^{2}}{4g^{2}\Omega^{3}}\right)}
\eeq
which coincide, exactly, with the solution given by Dinh Dang in the case where he
neglects the q-term. We see that in this case the Goldstone theorem is violated and 
therefore the particle number symmetry is not restored.
  
Dinh Dang also treats within the Fermion formalism several other approximations which do 
not produce the Goldstone mode at zero energy. One does not very well understand the aim 
of these considerations, since any way it is well known~\cite{b5,b4,b3} that the Goldstone 
mode should come at zero energy to restore the particle number symmetry which otherwise is 
violated. 

As a last point we would like to mention that Dinh Dang is superposing addition and
removal modes in the non superfluid phase in Eq.~(75). This superposition couples, 
simultaneously, $N\pm 2$ states and therefore, obviously, violates particle number conservation
already in the non superfluid phase. On the other hand it is well known~\cite{b5, b4, b3} that
pp(hh)-RPA in the non superfluid phase perfectly respects particle number symmetry :
the addition mode involves amplitudes $\langle N+2|a^{\dagger}a^{\dagger}|N\rangle$
and the removal mode $\langle N-2|aa|N\rangle$. None of these amplitudes violates
particle number.

In short, in this comment we pointed out a number of shortcomings and inconsistencies
in the work by Dinh Dang~\cite{b1} involving pp-RPA and QRPA in a solvable model.  
\begin{acknowledgments}
I would like to think Guy Chanfray, Jorge Dukelsky and Peter Schuck for discussions. 
\end{acknowledgments}

\end{document}